\documentclass[paper,showpacs,preprintnumbers,amsmath,amssymb]{revtex4}

\usepackage{graphicx}
\usepackage{dcolumn}
\usepackage{bm}

\begin{document}


\title{The scaling properties of dissipation in incompressible isotropic 
three-dimensional magnetohydrodynamic turbulence}

\author{J. A. Merrifield}
 \affiliation{Department of Physics, University of Warwick, Coventry CV4 7AL, UK}
\author{W.-C. M\"{u}ller}%
 \affiliation{Max-Planck-Institut f\"{u}r Plasmaphysik, 85748 Garching Germany}
\author{S. C. Chapman}%
 \affiliation{Department of Physics, University of Warwick, Coventry, UK}
 \altaffiliation[Also at ]{Radcliffe Institute, Harvard University, Cambridge, MA, USA}
\author{R. O. Dendy}%
 \affiliation{UKAEA Culham Division, Culham Science Centre, Abingdon, Oxfordshire OX14 3DB, UK}
 \altaffiliation[Also at ]{Department of Physics, University of Warwick, Coventry CV4 7AL, UK}
\date{\today}
\begin{abstract}
The statistical properties of the dissipation process constrain the analysis of large scale numerical simulations of three dimensional incompressible magnetohydrodynamic (MHD) turbulence, such as those of Biskamp and M\"{u}ller [{\em Phys. Plasmas} \textbf{7}, 4889 (2000)]. The structure functions of the turbulent flow are expected to display statistical self-similarity, but the relatively low Reynolds numbers attainable by direct numerical simulation, combined with the finite size of the system, make this difficult to measure directly. However, it is known that extended self-similarity, which constrains the ratio of scaling exponents of structure functions of different orders, is well satisfied. This implies the extension of physical scaling arguments beyond the inertial range into the dissipation range. The present work  focuses on the scaling properties of the dissipation process itself. This provides an important consistency check in that we find that the ratio of dissipation structure function exponents is that predicted by the She and Leveque [{\em Phys. Rev. Lett} \textbf{72}, 336 (1994)] theory proposed by Biskamp and M\"{u}ller. This supplies further evidence that the cascade mechanism in three dimensional MHD turbulence is non-linear random eddy scrambling, with the level of intermittency determined by dissipation through the formation of current sheets.\\
{\em Copyright (2005) American Institute of Physics. This article may be downloaded for personal use only. Any other use requires prior permission of the author and the American Institute of Physics.}
\end{abstract}
\maketitle
\section{Introduction} 
\begin{figure}
\begin{center}
\includegraphics[width=0.5\textwidth]{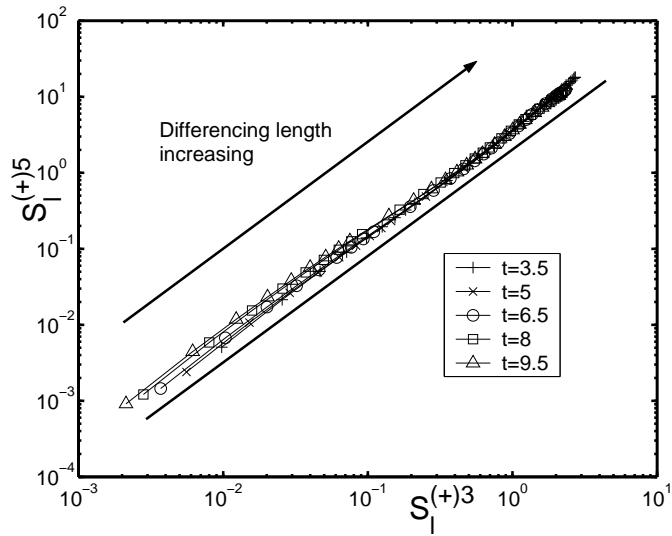}
\end{center} 
\caption{Extended self-similarity for the Els\"{a}sser field variable $\mathbf{z}^{(+)}$ (order five against order three), compare Eq.(\ref{ess}), for decaying MHD turbulence 
where structure functions are normalised by the total energy before time averaging. This normalisation reveals the same underlying scaling for points from different simulation 
times, as shown. After Biskamp and M\"{u}ller \cite{BiskMullPoP}.} 
\label{zp5_v_zp3}
\end{figure}

\begin{figure}
\begin{center}
\includegraphics[width=0.48\textwidth]{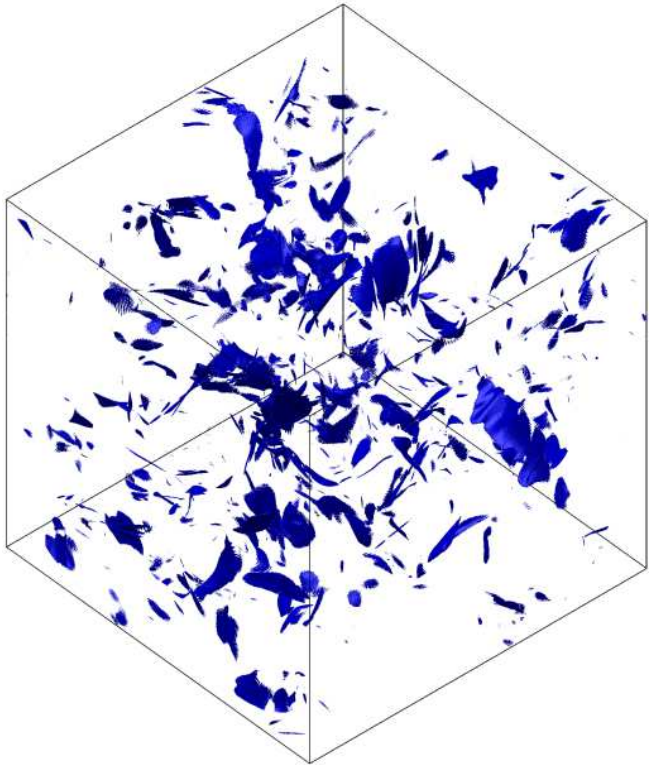}
\end{center}
\caption{(Color online) Isosurfaces of sheet-like (2D) coherent structures of the squared gradient of the $\mathbf{z}^{(+)}$ Els\"{a}sser field variable from the 3D MHD turbulence simulation ofBiskamp and M\"{u}ller.}
\label{isozp}
\end{figure}

This paper investigates the previously under explored topic of scaling in the local rate of dissipation in magnetohydrodynamic (MHD) flows. Turbulent fluids and plasmas display three properties that motivate development of statistical theories \cite{frisch}: (i) disorganisation, in the sense that structures arise on all scales; (ii) unpredictability of detailed behaviour, in the sense of inability to predict a signal's future behaviour from knowledge of its past, implying links with deterministic chaos; and (iii) reproducibility of statistical measures, combined with the presence of statistical self-similarity. Much progress has been made by the heuristic treatment of scaling laws derived from energy cascade arguments, following Kolmogorov and Richardson, see for example Ref.\cite{frisch}. The basic idea is that energy-carrying structures (eddies) are injected on large scales, non-linear eddy interaction causes energy to cascade to smaller scales in a self-similar manner, and energy is finally dissipated by viscosity on small scales. A quasi-stationary state evolves where the rate of viscous dissipation matches the rate of energy injection. Scaling exponents $\zeta_p$ characterise the resulting statistical self-similarity found in structure functions $S_l^p$:
\begin{equation} 
\label{str_fnc} 
S_l^p = \langle \left(\mathbf{v}(\mathbf{x}+\mathbf{l},t) \cdot \mathbf{l}/l-\mathbf{v}(\mathbf{x},t) \cdot \mathbf{l}/l\right)^p \rangle \sim l^{\zeta_p} 
\end{equation}
Here $\mathbf{v}$ is the fluid velocity, $\mathbf{x}$ is a position vector, $\mathbf{l}$ is a differencing vector, and the average is an ensemble average. The statistical 
self-similarity represented by the power-law in $l$ is only valid within the inertial range {$l_d \ll l \ll l_0$}; here $l_0$ is the characteristic macroscale, and $l_d$ is the dissipation scale at which the cascade terminates. The set of scaling exponents $\zeta_p$ in Eq.(\ref{str_fnc}) is expected to be universal since it characterises the generic cascade process. It is worth noting here that universality can only be expected in the isotropic case. When anisotropies are present, deviation from the isotropic case can be expected, and this will relate to the strength of the anisotropy. In MHD turbulence, anisotropy can be introduced in the form of an imposed magnetic field. The effect of this on the scaling exponents is investigated in Ref.\cite{MullerBiskampGrappin}. This reference also investigates anisotropy in terms of that introduced by the local magnetic field even when an applied field in absent. This stems from the Goldreich and Sridhar objection to the assumption of local isotropy in the MHD case \cite{GS3}. In Ref.\cite{MullerBiskampGrappin} structure functions are calculated with the differencing length perpendicular and parallel to the local magnetic field. The perpendicular structure functions were found to exhibit stronger intermittency than the parallel structure functions. Exponents calculated from the perpendicular structure functions were found to coincide with those calculated from the isotropic structure functions.  Essentially dimensional arguments, in which the relevant physical parameters are identified heuristically, have been formulated to provide basic fluid scaling information. These arguments linearly relate $\zeta_p$ to $p$, for example the Kolmogorov 1941 phenomenology \cite{K41,ref_sym} predicts $\zeta_p = p/3$. As such, basic fluid scaling can be characterised by one number $a$ such that
\begin{equation}
S_l^p \sim l^{pa}
\label{basic_scaling}
\end{equation}
To exploit these concepts, let us write the equations of incompressible MHD in Els\"{a}sser symmetric form \cite{BisBook}:
\begin{equation} 
\partial_t \mathbf{z}^\pm = -\mathbf{z}^\mp.\mathbf{\nabla z}^\pm - \mathbf{\nabla} \left(p+ B^2/2 \right) + \left(\nu/2 + \eta/2\right) \nabla^2 \mathbf{z}^\pm + \left(\nu/2 -
 \eta/2\right) \nabla^2 \mathbf{z}^\mp
\label{Elsasser_sym}
\end{equation}
\begin{equation}
\mathbf{\nabla}.\mathbf{z}^\pm = 0
\label{sol_cond}
\end{equation}
Here the Els\"{a}sser field variables are  $\mathbf{z}^\pm = \mathbf{v} \pm \mathbf{B} \left( \mu_0\rho\right)^{-\frac{1}{2}}$, where $p$ is the scalar pressure, $\nu$ is kinematic viscosity, $\eta$ is magnetic diffusivity and $\rho$ is fluid density. The symmetry of Eq.(\ref{Elsasser_sym}) suggests that statistical treatment of $\mathbf{z}^{\pm}$ may be more fundamental than separate treatments of $\mathbf{v}$ and $\mathbf{B}$. In light of this, longitudinal structure functions are constructed in terms of Els\"{a}sser field variables hereafter:
\begin{equation}
\label{Elsasser_strfnc} 
S_l^{p(\pm)} = \langle |\left(\mathbf{z}^{(\pm)}(\mathbf{x}+\mathbf{l},t) \cdot \mathbf{l}/l -\mathbf{z}^{(\pm)}(\mathbf{x},t) \cdot \mathbf{l}/l \right)|^p \rangle\sim l^{\zeta^{(\pm)}_p} 
\end{equation}
As mentioned above, heuristic arguments that make predictions about basic fluid scaling only linearly relate $\zeta_p$ to $p$. In reality $\zeta_p$ depends nonlinearly on $p$ due to the intermittent spatial distribution of eddy activity. Basic fluid scaling can be modified to take
this into account by the application of an intermittency correction. A commonly applied class of intermittency correction describes statistical 
self-similarity in the local rate of dissipation $\epsilon_l$ by means of 
scaling exponents $\tau_p$:  
\begin{equation}
\label{eps_strfnc}
\langle \epsilon_l^p \rangle  \equiv \langle \left(\frac{\nu}{4\pi l^3}\int_0^l\frac{1}{2}\left(\partial_iv_j(x+l',t)+\partial_jv_i(x+l',t)\right)^2dl'^3\right)^p\rangle \sim 
l^{\tau_p} 
\end{equation}
For a review of the fractal nature of the local rate of dissipation for hydrodynamics, see for example Ref.\cite{Men}. 
\\\\
As we shall see, the intermittent nature of the system is captured by the nonlinear dependence of $\tau_p$ on $p$ in Eq.(\ref{eps_strfnc}). This nonlinearity can conveniently be expressed multiplicatively in relation to the basic linear fluid scaling of Eq.(\ref{basic_scaling}). Specifically we may write
\begin{equation}
S_l^p \sim \langle \epsilon_l^{gp} \rangle l^{a p}
\label{gen_ref_sym}
\end{equation}
Where $g$ is a constant whose value is inferred from model assumptions such as those of Kolmogorov (K41) and Iroshnikov-Kraichnan (IK) \cite{Irosh,Kraich}. This is Kolmogorov's refined similarity hypothesis \cite{ref_sym}. The scaling exponents $\zeta_p$ in Eq.(\ref{Elsasser_strfnc}) are inferred by Eq.(\ref{gen_ref_sym}) to be $\zeta_p = \tau_{pg} + pa$. That is, the intermittency in the velocity field structure functions is achieved via reasoning concerning local rate of dissipation. One model that uses this hypothesis, and has proven successful in predicting the scaling exponents for hydrodynamic turbulence, is that from the 1994 paper of She and Leveque (SL) \cite{SL94}. Here physical assumptions are made regarding the scaling of the local rate of dissipation. Specifically: the hierarchical nature of Eq.(\ref{eps_strfnc}) above, as expressed in Eq.(6) of Ref.\cite{SL94}; the rate of dissipation by the most intensely dissipating structures is related only to the eddy turnover time as determined by the basic fluid scaling, as in Eq.(4) of Ref.\cite{SL94}; and the space filling nature of the most intensely dissipating structures can be described by one parameter (their Hausdorff dimension). These three assumptions can be combined to formulate a second order difference equation for the scaling exponents $\tau_p$ that has one non-trivial solution, as in Eq.(9) of Ref.\cite{SL94}. This solution can be formulated in terms of the two parameters: the co-dimension of the most intensely dissipating structures, $C = D - d_H$, where $D$ is the embedding dimension and $d_H$ is the Hausdorff dimension; and the basic fluid scaling number expressed by $a$ in Eq.(\ref{basic_scaling}) above. Following Ref.\cite{SL94}, we may write
\begin{equation}
\tau_p = -(1-a) p + C - C(1-(1-a)/C)^p
\label{gen_tau_p}
\end{equation}
This two parameter formulation follows that previously noted by Dubrulle \cite{Dubrulle}, whose parameters $\Delta$ and $\beta$ correspond to our $(1-a)$ and $1-\Delta/C$ respectively. The refined similarity hypothesis, as expressed in Eq.({\ref{gen_ref_sym}), is then invoked to obtain the following expression for the structure function scaling exponents $\zeta_p$:}
\begin{equation}
\zeta_p = pa -(1-a)pg + C - C(1-(1-a)/C)^{pg}
\label{gen_zeta_p}
\end{equation}
Previously Els\"{a}sser field structure functions have been identified with an SL model of the type Eq.(\ref{gen_zeta_p}), see, for example Refs.\cite{BiskMullPoP,Axel}. In the present paper the refined similarity hypothesis for MHD is tested by applying a modified form of Eq.(\ref{gen_ref_sym}), see Eq.(\ref{mod_k_ref}), to the simulation data of Biskamp and M\"{u}ller. This provides an important consistency check for previous studies. Equation (\ref{gen_tau_p}), which probes the multifractal properties of the local rate of dissipation, but does not rely on the refined similarity hypothesis, can also be tested directly against the simulation results, as we discuss below.
\\\\
Direct numerical simulations must resolve the dissipation scale $l_d$ so that energy does not accumulate at large wavenumbers, artificially stunting the cascade. Most of the numerical resolution is therefore used on the dissipation range, whereas it is only on scales much larger than $l_d$ that dissipative effects are negligible, and scaling laws of the type discussed arise. Thus high Reynolds number simulations with an extensive inertial range are currently unavailable. However, the principle of extended self-similarity (ESS) \cite{Benz_ESS} can be used to extend the inertial range scaling laws into the range of length scales that is significantly affected by dissipation but still larger than $l_d$.  Instead of considering the scaling of individual structure functions, the  principle of ESS involves investigating the scaling of one order structure function against another, on the assumption that  
\begin{equation} 
\label{ess} 
S_l^{p(\pm)} \sim \left(S_l^{q(\pm)}\right)^{ \left( \zeta_p/\zeta_q \right)} 
\end{equation}
Here it can be seen that any set of structure functions will satisfy this relation providing
\begin{equation}
S_l^p \sim G(l)^{\zeta_p}
\label{ess_recast}
\end{equation}
where $G(l)$ can be any function of $l$ which is independent of $p$. Here we use the notation of S. C. Emily {\em et al.} that these authors used to describe the general properties of generalised extended self-similarity, as expressed in Eq.(8) of Ref.\cite{Emily} -- though generalised ESS is not discussed in the present paper. When the objective of ESS is to extend scaling into the dissipation range, $G(l)$ can be rewritten as $l^{G'(l)}$, where $G'(l)$ is introduced to accommodate the non constant flux of energy through length scales in the dissipation range. As such, $G'(l)$ asymptotically approaches one as the length scale increases from the dissipative to the inertial range.
\\\\
The She-Leveque model as it has appeared so far, is only valid in the inertial range. Let us now discuss how this model can be interpreted in the framework of ESS. This problem has been tackled for hydrodynamic turbulence by Dubrulle \cite{Dubrulle} for example. In that paper the explicit inclusion of $l$ in the refined similarity hypothesis [Eq.(\ref{gen_ref_sym}) with $g=a=1/3$ for hydrodynamic turbulence] is replaced by a generalised length scale, which is cast in terms of the third order structure function as expressed in Eq.(12) of Ref.\cite{Dubrulle}. This problem was addressed similarly by Benzi {\em et al.} where the scaling relation 
\begin{equation}
\label{mod_k_ref} 
S_l^p \sim \langle \epsilon^{gp}_l \rangle\left( S_l^{1/a} \right)^{pa} 
\end{equation} 
is explicitly formulated in Ref.\cite{Benz_ESS2}. The appropriate relation between $\zeta_p$ and $\tau_p$ is now $\zeta_{p} = \tau_{pg} + pa\zeta_{1/a}$. Using this relation combined with Eq.(\ref{ess_recast}) and Eq.(\ref{mod_k_ref}) it can be seen that $\langle \epsilon_l^p \rangle$ must also have the form $\langle \epsilon_l^p \rangle \sim G(l)^{\tau_p}$. This implies ESS exists also in the local rate of dissipation, such that
\begin{equation}
\label{ess_epsilon} 
\langle \epsilon_l^p \rangle \sim \langle \epsilon_l^q \rangle ^{\left( \tau_p / \tau_q \right)} 
\end{equation} 
It can then be seen that if a She-Leveque model of the general type Eq.(\ref{gen_zeta_p}) is used to explain scaling exponents obtained via ESS, as expressed in Eq.(\ref{ess}), then two consistency checks are appropriate. First Kolmogorov's refined similarity hypothesis should be satisfied in the form Eq.(\ref{mod_k_ref}), and second ESS should exist in the local rate of dissipation as in Eq.(\ref{ess_epsilon}).
\\\\
The present paper performs these checks for consistency for the simulation of Biskamp and M\"{u}ller \cite{BiskMullPoP}. Here the scaling exponents $\zeta^{(\pm)}$ [see Eq.(\ref{Elsasser_strfnc})] were investigated via direct numerical simulation of the three dimensional (3D) incompressible MHD equations, with a spatial grid of $512^3$ points \cite{BiskMullPoP,BiskMullPRL}. The simulation is of decaying turbulence with initially equal magnetic and kinetic energy densities and $\nu$ = $\eta$. A fuller discussion of the numerical procedure is present in the next section. Since the turbulence decays with time, structure functions are normalised by the total energy in the simulation (kinetic plus magnetic) before time averaging takes place. Biskamp and M\"{u}ller \cite{BiskMullPoP} extracted the ratios of scaling exponents $\zeta_p / \zeta_3$ by ESS and directly determined $\zeta_p \simeq 1$. These exponents were found to match a variant of the She-Leveque 1994 model Eq.(\ref{gen_zeta_p}) inserting Kolmogorov basic fluid scaling ($g = a = 1/3$) with the most intensely dissipating structures being sheet-like.($C = 1$). Early investigations of this type assumed Iroshnikov-Kraichnan fluid scaling where the most intensely dissipating structures are sheet-like \cite{pol_pouq,grauer_krug:mhdsl} (see Ref. \cite{pol_pouq} for a derivation of $\zeta_p^{(\pm)}$ for this case). Sheet-like intensely dissipating structures are thought to exist in MHD turbulence because of the propensity of current sheets to form. We refer to Fig.5 of Ref.\cite{BiskMullPoP} for isosurfaces of current density squared, and to Fig.\ref{isozp} for isosurfaces constructed from the shear in the $\mathbf{z}^{(+)}$ field $\left( \partial_i z^{(+)}_i \right)^2$. Both figures show the existence of two dimensional coherent structures; Fig.\ref{isozp} is more directly related to the analyses presented in the present paper. Basic Kolmogorov fluid scaling for Alfv\'{e}nic fluctuations has been verified for low Mach number ($\simeq 0.1$) compressible \cite{Axel,Boldyrev} and incompressible \cite{BiskMullPoP,BiskMullPRL} 3D MHD turbulence by power spectrum analysis, and by checking for scaling in the third order structure function such that $\zeta_3 = 1$. Extended self-similarity has also been utilised to extract ratios of scaling exponents related to an inverse cascade in laboratory plasmas \cite{Antar}. In other work, a generalised version of this SL model has been applied to compressible flows where $C$ is allowed to vary as a fitting parameter \cite{Axel,Padoan}, and in the case of Ref.\cite{Padoan} this dimension is interpreted as a function of the sonic Mach number. Figure \ref{zp5_v_zp3} shows an example of the normalisation and ESS procedure for $\mathbf{z}^{(+)}$ structure functions from the data analysed here.
\\\\
The philosophy behind our investigation can now be summarised as follows. Given a simulation, the set of structure functions $S_l^p$ can be calculated. These are expected to display statistical self-similarity as  expressed in Eq.(\ref{Elsasser_strfnc}), where the scaling exponents $\zeta_p$ give insight into the physics of the cascade process. The relatively low Reynolds numbers attainable by direct numerical simulation, combined with the finite size of the system, make this statistical self-similarity difficult to measure directly. However, it is found that extended self-similarity of the type expressed in Eq.(\ref{ess}) is well satisfied, allowing the ratio of scaling exponents $\zeta_p/\zeta_3$ to be directly measured. There is a range of {\em a priori} views concerning these ratios, reflecting physical model assumptions. The ratios of scaling exponents recovered from ESS analysis of the 3D MHD simulation data are compared with these models, and the best fit is identified. Our investigation thus assists in validating the physical assumptions made in formulating the currently favoured model, namely Eq.(\ref{gen_zeta_p} with $g=a=1/3$ and $C=1$  giving $\zeta_p=p/9+1-(1/3)^{p/3}$. In particular, we confirm the existence of a specific type of extended self-similarity in the local rate of dissipation, with exponents given by Eq.(\ref{gen_tau_p}) with $a=1/3$ and $C=1$ giving $\tau_p = -2p/3 + 1 - (1/3)^p$. We also show that Kolmogorov's refined similarity hypothesis, in the form Eq.(\ref{mod_k_ref}), is satisfied. 

\section{Numerical Procedures}
The data analysed here stems from a direct numerical simulation of decaying isotropic turbulence (see Ref.\cite{BiskMullPoP} for additional details). The equations of incompressible MHD are written in terms of the vorticity, $\bf{\omega}=\mathbf{\nabla}\times\mathbf{v}$, in order to eliminate the pressure variable. These are solved by a pseudospectral scheme (see, for example, Ref.\cite{canuto:book}). Time integration is carried out by applying the trapezoidal leapfrog method 
\cite{kurihara:trapezleapfrog}. The aliasing error associated with this approach \cite{orszag:aliasing} is reduced to discretisation error level by spherical truncation of the 
Fourier grid \cite{vincent_meneguzzi:simul}. 
\\\\
The simulation domain comprises a periodic box in Fourier space with $512^3$ points. Initially the fields have random phases and amplitudes 
$\sim\exp(-k^2/(2k_0^2))$ with $k_0=4$.  The ratio of total kinetic and magnetic energy of the fluctuations is set to one. Cross helicity, which is proportional to $\int_V \left( \left(\mathbf{z}^+\right)^2-\left(\mathbf{z}^-\right)^2 \right) dV$, is absent throughout the duration of the simulation. The magnetic helicity, $H^\mathbf{M}=\frac{1}{2}\int_V \mathbf{A}\cdot\mathbf{B} dV$, where  $\mathbf{B}=\nabla\times\mathbf{A}$, is set to 
$0.7 H^\mathrm{M}_{\max}$. Here $H^\mathbf{M}_{\max}\simeq E^M/k_0$ where $E^M$ is the energy in the magnetic field. The diffusivities $\nu=\eta=4\times10^{-4}$ imply a magnetic Prandtl number $Pr_m=\nu/\eta$ of one. 
\\\\
The run was performed over ten eddy turnover times, defined as the time required to reach maximum dissipation when starting from smooth initial fields. Structure functions and moments of dissipation are calculated in time intervals of $0.5$ between $t=3.5$ and $t=9.5$, during which the turbulence undergoes self-similar decay.

\section{Results}
\begin{figure}
\begin{center}
\includegraphics[width=0.5\textwidth]{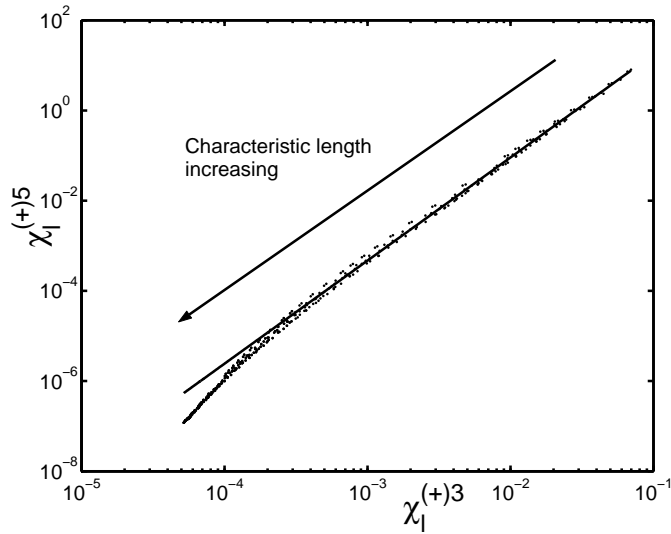}
\end{center}
\caption{Extended self-similarity in the Els\"{a}sser field variable $\mathbf{z}^{(+)}$ gradient squared proxy for the local rate of dissipation (order five against order three), compare 
Eq.(\ref{ess_epsilon}) with the gradient squared proxy from Eq.(\ref{grad_square}) replacing $\epsilon_l^p$.  Normalisation by the space averaged local rate of viscous and 
Ohmic dissipation allows time averaging in spite of the decay process. Deviation from power law scaling at large $l$ is probably a finite size effect. The solid line is the 
best fit in the linear region.}
\label{chi5_v_chi3}
\end{figure}

\begin{figure}
\begin{center}
\includegraphics[width=0.5\textwidth]{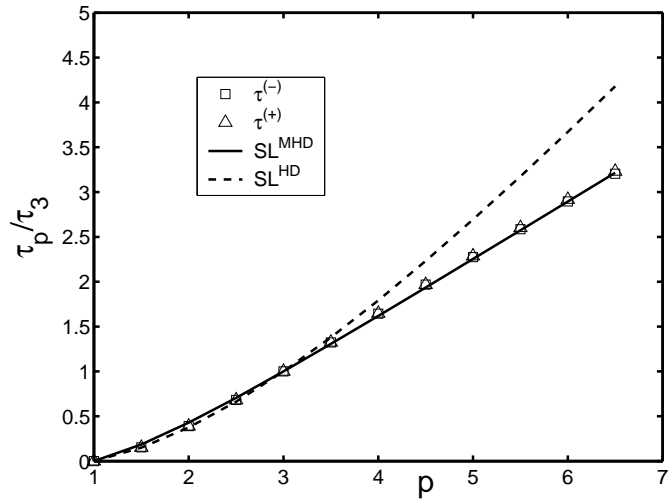}
\end{center}
\caption{Ratio of scaling exponents (order p over order three) obtained via extended self-similarity from the Els\"{a}sser field gradient squared proxy for the local rate of dissipation. Errors in these measurements lie within the marker symbols. Solid line shows ratios predicted by a She-Leveque theory based on Kolmogorov fluid scaling and sheet-like most intensely dissipating structures, Eq.(\ref{SL_MHD}). The dashed line shows ratios predicted by hydrodynamic She-Leveque \cite{SL94}.}
\label{taup_v_p}
\end{figure}

\begin{figure}
\begin{center}
\includegraphics[width=0.5\textwidth]{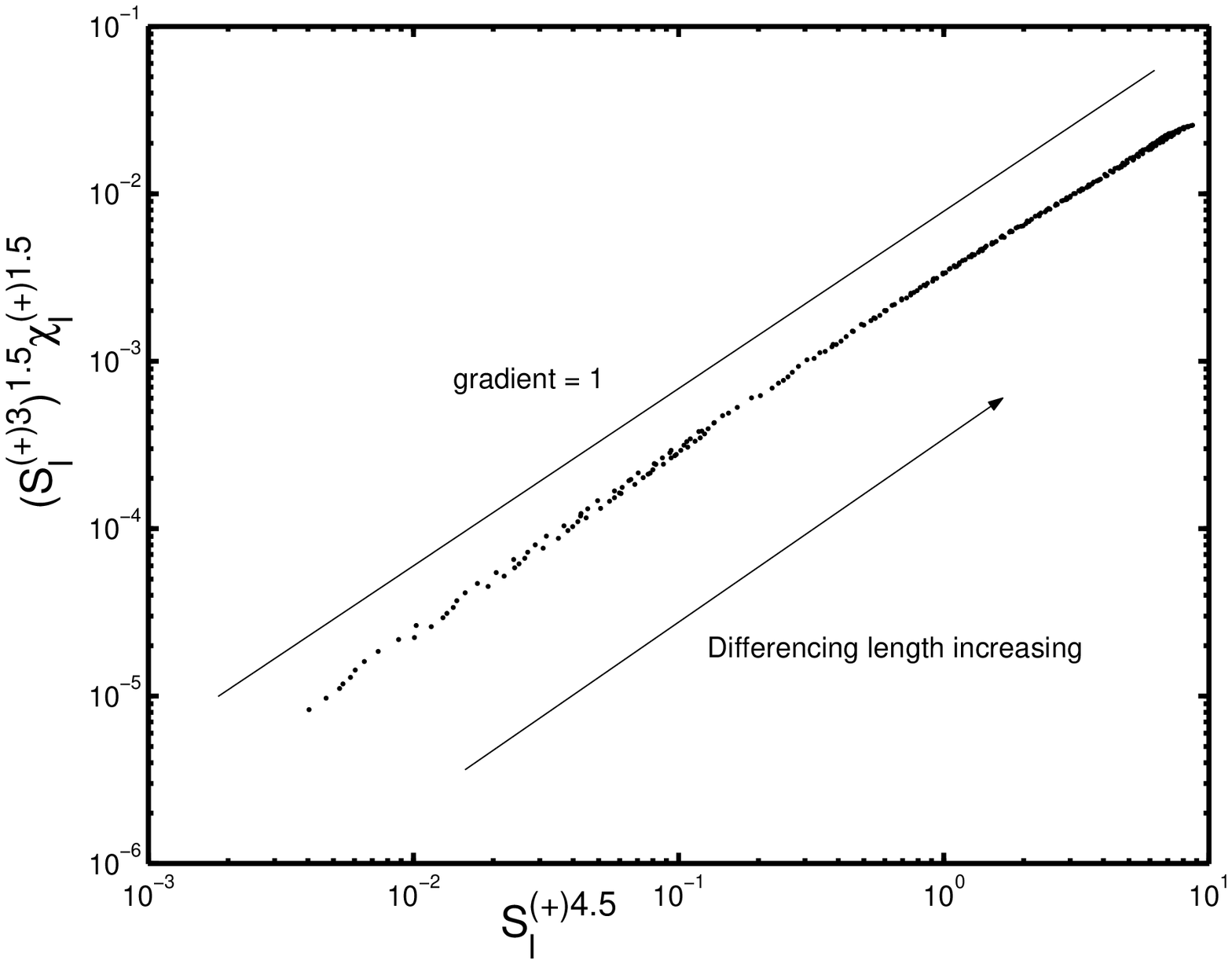}
\end{center} 
\caption{Plot to test Kolmogorov's refined similarity hypothesis as applied to extended self-similarity, Eq.(\ref{mod_k_ref}). This involves taking the product of the field variable and dissipative structure functions, in contrast to Figs.\ref{zp5_v_zp3} and \ref{chi5_v_chi3}. Agreement with the hypothesis would give a straight line with unit gradient. Normalisation was performed as in Figs.\ref{zp5_v_zp3} and \ref{chi5_v_chi3} to allow time averaging despite the decay process.}
\label{cross_plot_n1p5} 
\end{figure}

\begin{figure}
\begin{center}
\includegraphics[width=0.5\textwidth]{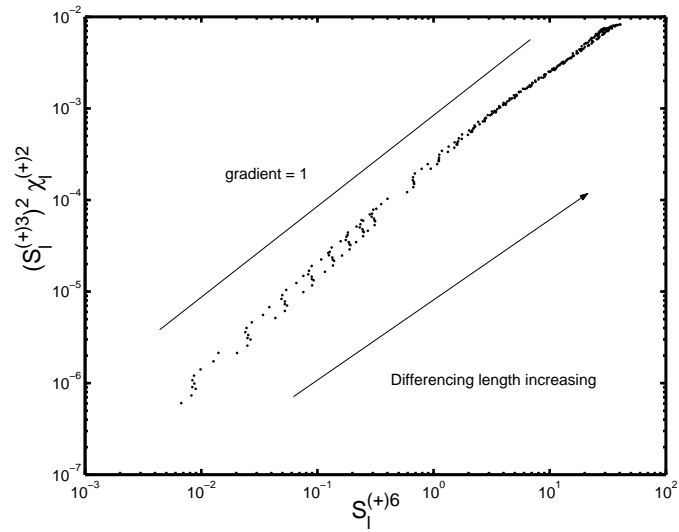}
\end{center} 
\caption{High order test of Kolmogorov's refined similarity hypothesis as applied to extended self-similarity, Eq.(\ref{mod_k_ref}). Normalisation is performed as in Fig.\ref{cross_plot_n1p5}.}
\label{cross_plot_n2}  
\end{figure}

In the present paper, the gradient squared measure $(\partial_iz_i^{(\pm)})^2$ is used as a proxy \cite{ref_sym} for the local rate of dissipation $\left(\partial_i{B}_j-\partial_j{B}_i\right)^2\eta/2+\left(\partial_i{v}_j+\partial_j{v}_i\right)^2\nu/2$. This proxy has recently been employed to study turbulence in the thermal ion velocity of the solar wind as measured by the ACE spacecraft \cite{Bersh}, giving results consistent with those presented below. This is particularly interesting insofar as the solar wind remains one of the few accessible system of quasistationary fully developed MHD turbulence \cite{BisBook} although we note that MHD intermittency studies have also been performed on the reversed field pinch experiment RFX \cite{RFXCarbone}. Figure \ref{isozp} shows isosurface plots of the gradient squared measure for the simulation of Biskamp and M\"{u}ller. Two dimensional coherent structures dominate the image, suggesting the dimension parameter entering an SL model should equal two, as in the model employed by Biskamp and M\"{u}ller. Following Eq.(\ref{eps_strfnc}), statistical self-similarity in the dissipation measure is expressed as 
\begin{equation} 
\label{grad_square} 
\chi_l^{p(\pm)} \equiv \langle \left( \frac{1}{l} \int_0^l \left( \partial_i{z}^{(\pm)}_i({x}+{l}',t) \right)^2dl' \right)^p \rangle \sim l^{\tau_p^{(\pm)}}
\end{equation} 
This proxy, which involves a one dimensional integration rather than the full 3D integration of Eq.(\ref{eps_strfnc}), facilitates comparison with related experimental MHD \cite{Bersh,Antar} and hydrodynamic \cite{Benz_ESS2,Men,Chavarria} studies, and also offers benefits in computation time. 
\\\\
The SL model adopted by Biskamp and M\"{u}ller predicts
\begin{equation} 
\tau_p^{(\pm)} = - 2p/3 + 1 - \left(1/3\right)^p 
\label{SL_MHD}
\end{equation}
This is simply Eq.(\ref{gen_tau_p}) with $a = 1/3$ and $C = 1$. Gradients are calculated from the data of Biskamp and M\"{u}ller \cite{BiskMullPoP} using a high order finite difference scheme, and the integral is performed by the trapezium method. Normalisation by the spatial average of viscous plus Ohmic rates of dissipation allows time averaging to be performed. Figure \ref{chi5_v_chi3} shows an example of the ESS and normalisation procedure for $\chi_l^{p(+)}$ order $p=5$ against order $p=3$.  Statistical self-similarity is recovered with roll-off from power law scaling as $l$ aproaches the system size. This roll-off behaviour at large $l$ may be due to the finite size of the system, since a more extensive part of the simulation domain is encompassed by the spatial average (the integral over $dl'$) as $l$ increases in Eq.(\ref{grad_square}). In Fig.\ref{chi5_v_chi3} points identified with this roll-off are removed, and ratio of scaling exponents $(\tau_p/\tau_3)$ is calculated from the remaining points by linear regression. These ratios are shown in Fig.\ref{taup_v_p}. No significent difference between the scaling recovered from $\mathbf{z}^{(+)}$ and $\mathbf{z}^{(-)}$ can be seen. This should be expected since no theoretical distinction needs to be drawn between $\mathbf{z}^{(+)}$ and $\mathbf{z}^{(-)}$ for the vanishing values of cross helicity present in this simulation. The solid line in Fig.\ref{taup_v_p} shows the ratio predicted by Eq.(\ref{SL_MHD}), in contrast to the dashed line which shows the ratio predicted by the SL theory for hydrodynamic turbulence \cite{SL94}. Caution must be taken when calculating high order moments, since these are strongly influenced by the tails of their associated distributions.  This can easily lead to bad counting statistics. The order $p$ is only taken up to $p=6.5$ for the dissipation measure (instead of $p=8$ as for the Els\"{a}sser field structure functions \cite{BiskMullPoP}) because of the extremely intermittent nature of the signal; large values affect the average [the angular brackets in Eq.(\ref{grad_square})] more as the order $p$ increases. This effect is evaluated using a similar methodology to that in Ref.\cite{NPG}. If a worst case scenario is imagined, where the average is strongly affected by one point in the signal, one would expect $l / \delta l$ members of the spatial average in Eq.(\ref{grad_square}) to be directly affected by this point, where $\delta l$ is the grid spacing. We can then define an event as incorporating $l / \delta l$ members of the  spatial average. It is found that $\simeq5$ percent of the average is defined by only $\simeq10$ events for order $p=6.5$. This situation is of course worse for higher values of $p$.
\\\\
Plots were constructed in order to test Eq.(\ref{mod_k_ref}). This involves taking the product of structure functions of the field variables and the dissipative quantities, in contrast to Figs.\ref{zp5_v_zp3} and \ref{chi5_v_chi3}. Figures \ref{cross_plot_n1p5} and \ref{cross_plot_n2} show these plots for $n=1.5$ and $n=2$ respectively. The low order measure in Fig.\ref{cross_plot_n1p5} shows a relation that is nearly linear, with a gradient close to the ideal value of one, see Eq.(\ref{mod_k_ref}). This is encouraging considering the deviation expected at the smallest and largest scales due to finite size effects. However, unlike the case in Fig.\ref{chi5_v_chi3}, there may be some curvature across the range of the plot. The higher order measure in Fig.\ref{cross_plot_n2} deviates from a linear scaling relation. We note that taking this test to high order involves the product of two quantities that have challenging counting statistics, plotted against a high order structure function. The deviation of the gradient seen in Fig.\ref{cross_plot_n2} from the ideal value of one is perhaps not surprising, because the constraints described above become stronger at high order.
\section{Conclusions} 
Extended self-similarity is recovered in the gradient squared proxy for the local rate of dissipation of the Els\"{a}sser field variables $\mathbf{z}^{(\pm)}$ computed by Biskamp and M\"{u}ller. We believe this is the first time this has been shown for MHD flows. This result supports the application to Els\"{a}sser field scaling exponents $\zeta_p^{(\pm)}$ of turbulence theories that require statistical self-similarity in the local rate of dissipation, even when $\zeta_p^{(\pm)}$ are extracted from relatively low Reynolds number flows via ESS. Furthermore the ratio of exponents recovered is that predicted by the SL theory proposed by Biskamp and M\"{u}ller \cite{BiskMullPoP}. This supplies further evidence that the cascade mechanism in three dimensional MHD turbulence is non-linear random eddy scrambling, with the level of intermittency determined by dissipation through the formation of two dimensional coherent structures. However, Kolmogorov's ESS modified refined similarity hypothesis remains to be verified at high order. 
\\
\begin{acknowledgments}
We are grateful to Tony Arber for helpful discussions.
This research was supported in part by the United Kingdom Engineering and Physical Sciences Research Council. SCC acknowledges a Radcliffe fellowship.
\end{acknowledgments}
\clearpage
\bibliography{JMerrifPhysPlas14Sep}

\end{document}